\documentclass[twocolumn,showpacs,preprintnumbers,amsmath,nofootinbib,amssymb]{revtex4}
\input psfig.sty
\def \be {\begin{equation}}
\def \ee {\end{equation}}
\def \bea {\begin{eqnarray}}
\def \eea {\end{eqnarray}}

\usepackage{graphicx}% Include figure files
\usepackage{dcolumn}% Align table columns on decimal point
\usepackage{bm}% bold math

\begin{document}

\title{Hybrid dark energy}

\author{J. S. Alcaniz$^{1,2}$}
\email{alcaniz@on.br}

\author{R. Silva$^{3,4}$}
\email{raimundosilva@uern.br}

\author{F. C. Carvalho$^{5}$}
\email{fabiocc@das.inpe.br}

\author{Z.-H. Zhu$^{6}$} 
\email{zhuzh@bnu.edu.cn}

\affiliation{$^{1}$Observat\'orio Nacional, 20921-400 Rio de Janeiro - RJ, Brasil}

\affiliation{$^{2}$Instituto Nacional de Pesquisas Espaciais/CRN, 59076-740, Natal - RN, Brasil}

\affiliation{$^{3}$Universidade do Estado do Rio Grande do Norte,
59610-210, Mossor\'o - RN, Brasil}

\affiliation{$^{4}$Universidade Federal do Rio Grande do Norte, 59072-970 Natal - RN, Brasil}

\affiliation{$^{5}$Instituto Nacional de Pesquisas Espaciais, 12227-010, S\~ao Jos\'e dos Campos - SP, Brasil}

\affiliation{$^{6}$Department of Astronomy, Beijing Normal University, Beijing 100875, China}


\date{\today}

\begin{abstract}
Extending previous results [Phys. Rev. Lett. 97, 081301 (2006)], we explore the cosmological implications of a new quintessence scenario driven by a slow rolling homogeneous scalar field whose equation of state behaved as freezing over the entire cosmic evolution, is approaching $-1$ today, but will become thawing in the near future, thereby driving the Universe to an eternal deceleration. We argue that such a mixed behavior, named \emph{hybrid}, may reconcile the slight preference of current observational data for freezing potentials with the impossibility of defining observables in the String/M-theory context due to the existence of a cosmological event horizon in asymptotically de Sitter universes as, e.g., pure freezing scenarios.

\end{abstract}

\maketitle

\emph{Introduction}---Astronomical observations including distance measurements to intermediary and high-$z$ Type Ia Supernovae (SNe Ia) \cite{davis,wv07,Astier06,Riess07}, measurements of the Cosmic Microwave Background (CMB) anisotropies \cite{Sperg07}, and the current observations of the Large-Scale Structure (LSS) in the Universe \cite{tegmark04,bao} seem to indicate consistently that we live in a nearly flat, accelerating universe composed of $\sim 25\%$ of pressureless matter (baryonic + dark) and $\sim$ 75\% of an exotic component with large negative pressure, the so-called dark energy. In particular, this component is theoretically identified by its equation-of-state (EoS) parameter $w=p/\rho$, i.e., the ratio of the pressure $p$ to its energy density $\rho$. The simplest and most theoretically appealing candidate for dark energy is the constant cosmological $\Lambda$, for which $w=-1$. Another possibility is a dynamical field $\Phi$, the so-called quintessence field, whose the spatially-averaged EoS $-1 \leq w \leq -1/3$ is a time-dependent quantity (see \cite{revde} for recent reviews).

For this latter class of models, the physics behind the phenomenon of cosmic acceleration is the dynamics of a scalar field $\Phi$ rolling down its potential $V(\Phi)$, whose Lagrangian is simply given by ${\cal{L}} = {1\over2}\partial^{\mu}\Phi\partial_{\mu}\Phi-V(\Phi)$. In fact, this idea has received much attention over the past years and a considerable effort  has been made in understanding the role of quintessence fields on the dynamics of the Universe \cite{RatraPeebles,wetterich88,PRL95_141301,nos06}. Although from a physical viewpoint all these quintessence scenarios are based on the very same premise that fundamental physics provides motivation for light scalar fields in nature, in what concerns the dynamics and evolution of the Universe, it is well known that they may differ significantly among themselves.

In this regard, an interesting classification for the behavior of the quintessence field in terms of the time variation of its EoS ($w'= dw/d\ln a$) has been discussed in Refs. \cite{PRL95_141301,bs}. \emph{Thawing} models describe a scalar field whose EoS increases from $w \sim -1$, as it rolls down toward the minimum of its potential, whereas \emph{cooling} scenarios describe an initially $w > -1$ EoS decreasing to more negative values. A special case of this latter class of scenarios are the so-called \emph{freezing} models, which correspond to the situation in which the potential has a minimum at $\Phi = \infty$. 

Another possibility, not yet explored, would be the one in which the scalar field EoS behaved as freezing over all the past cosmic evolution, is approaching the value $-1$ today (in agreement with current observations), will become thawing in the near future and will behave as such  over the entire future evolution of the Universe. This mixed behavior is particularly interesting because, in principle, it could reconcile current observational and theoretical arguments about the dark energy EoS as, e.g., the slight preference of the SNe Ia + LSS data for freezing EoS \cite{krauss,trotta,huterer} (which in turn leads to an eternally accelerating Universe), and the impossibility of constructing a conventional S-matrix describing particle interactions in the String/M-theory context due to the existence of a cosmological event horizon in such freezing scenarios \cite{fischler}. 

In this \emph{Letter}, by following the arguments of Ref.~\cite{nos06}, and taking into account the theoretical and observational constraints on the behavior of the dark energy EoS above mentioned, we study a class of field potentials which gives origin to a mixed (freezing/thawing) EoS behavior, named here as \emph{hybrid}. We show that such a class of potentials may have a conventional dependence on the scalar field $\Phi$, i.e., $V(\Phi) \sim f_{\kappa}(\Phi) e^{g_{\kappa}(\Phi)}$ and may be obtained through a simple \emph{ansatz} on the scale factor derivative of the field energy density. Besides, it fully reproduces the exponential potential studied by Ratra and Peebles in Ref. \cite{RatraPeebles} for the dimensionless index $\kappa = 0$, and admits a wider range of solutions $\forall$  $\kappa \neq 0$. Due to the future thawing behavior ($w \rightarrow +1$ for $a \rightarrow \infty$), we also show that a transient accelerating phase is another interesting feature of this class of potentials, which in turn reconcile the observed acceleration of the Universe with the requirements of String/M theories discussed in Ref.~\cite{fischler}.

\emph{A Hybrid Model}---The action for the model is given by $S={m^2_{pl}\over 16\pi}\int d^4 x \sqrt{-g}\left[R -{1\over2}\partial^{\mu}\Phi\partial_{\mu}\Phi-V(\Phi)+{\cal{L}}_{m}\right]$, where $R$ is the Ricci scalar and $m_{pl}\equiv 1/\sqrt{G}$ is the Planck mass. The scalar field is assumed to be homogeneous, such that $\Phi=\Phi(t)$ and the Lagrangian density ${\cal{L}}_{m}$ includes all matter fields. The conservation equation for this $\Phi$ component takes the form $\dot\rho_{\Phi}+3H(\rho_{\Phi}+p_{\Phi})=0$, or, equivalently, $\ddot\Phi+3H\dot\Phi+V'(\Phi)=0\,,$ where $\rho_{\Phi}={1\over2}\dot\Phi^2+V(\Phi)$ and $p_{\Phi}={1\over2}\dot\Phi^2-V(\Phi) $ are, respectively, the field energy density and pressure. In the above expressions as well as in the subsequent ones, dots and primes denote, respectively, derivatives with respect to time and to the field (We work in units where $\hbar = c = 1$).

Following Ref.~\cite{nos06}, we will adopt here the following \emph{ansatz} on the scale factor derivative of the energy density
\be 
\label{ansatz}
\frac{1}{\rho_{\Phi}}\frac{\partial\rho_{\Phi}}{\partial a} = -A
\left(\frac{a^{\kappa-1/2}+a^{-\kappa-1/2}}{2}\right)^2\;, \ee
where $\kappa$ is a real parameter, $A$ is a positive number, and the other numeric factors were introduced for mathematical convenience. 

For a dark energy-dominated universe (${\cal{L}}_{m} = 0$), a direct combination of the above \emph{ansatz} with the conservation law for the field  and the Friedmann equation provides
\bea
\label{phi2}
\Phi(a) - \Phi_0 &=& {1\over\sqrt{\sigma}}\ln_{\kappa}(a)\;,
\eea
where $\Phi_0$ is the current value of the scalar field $\Phi$, $\sigma \equiv {8\pi/m_{\rm pl }^2 A}$, and the function $\ln_{\kappa}$ is a one-parameter generalized logarithmic function, defined as $\ln_{\kappa}(x)\equiv (x^{\kappa} - x^{-\kappa})/2\kappa$ with $|\kappa| \leq 1$, which reduces to the ordinary logarithmic as $\kappa \rightarrow 0$ \cite{kappa}. 

The potential $V(\Phi)$ for the above scenario is obtained by inserting the scale factor $a(\Phi)=\exp_{\kappa}[\sqrt{\sigma}(\Phi(a)-\Phi_0)]$ [from Eq.~(\ref{phi2})] into the above definitions of $\rho_{\Phi}$ and $p_{\Phi}$. The resulting potential is given by\footnote{Note that, in the inversion of the Eq.~(\ref{phi2}), we have used the one-parameter deformation of the exponential function 
$\exp_{_{\{{\scriptstyle \xi}\}}}(x)\equiv\left(\sqrt{1+\xi^2x^2}+\xi x\right)^{1/\xi}$, 
which not only reduces to an ordinary exponential in the limit $\xi\rightarrow0$ 
but also is the inverse function of the deformed logarithmic $\ln_{\xi}$.}
\bea
\label{gpotential}
V(\Phi) = f_{\kappa}(\Phi)
\exp\left[-\frac{A\sqrt{\sigma}\Phi
(1 + \kappa^2\sigma\Phi^2)g_{\kappa}(\Phi)}
{(\sqrt{1+\kappa^2\sigma\Phi^2} + \kappa\sqrt{\sigma}\Phi)^2}\right],
\eea
where 
\begin{equation}
f_{\kappa}(\Phi)= %\left[
1-\frac{A}{6}\left[\frac{1 - \kappa^2\sigma\Phi^2 + \kappa\sqrt{\sigma}\Phi g_{\kappa}(\Phi)}
{\sqrt{1+\kappa^2\sigma\Phi^2} + \kappa\sqrt{\sigma}\Phi}\right]^2,
\end{equation}
and $g_{\kappa}(\Phi) \equiv \sqrt{1+\kappa^2\sigma\Phi^2} + 2 \kappa\sqrt{\sigma}\Phi$. In the limit $\kappa \rightarrow 0$ Eqs.~(\ref{phi2}) and (\ref{gpotential}) reduce to $\Phi(a) - \Phi_0 = {1\over \sqrt{\sigma}}\ln(a)$ and $V(\Phi)\propto \exp(-A\sqrt{\sigma}\Phi)$, which fully reproduce the exponential potential studied in Ref.~\cite{RatraPeebles}. $\forall$ $\kappa \neq 0$, however, the scenario described above represents a generalized model which admits a much wider range of solutions, as described below.

\begin{figure*}[t]
\centerline{\psfig{figure=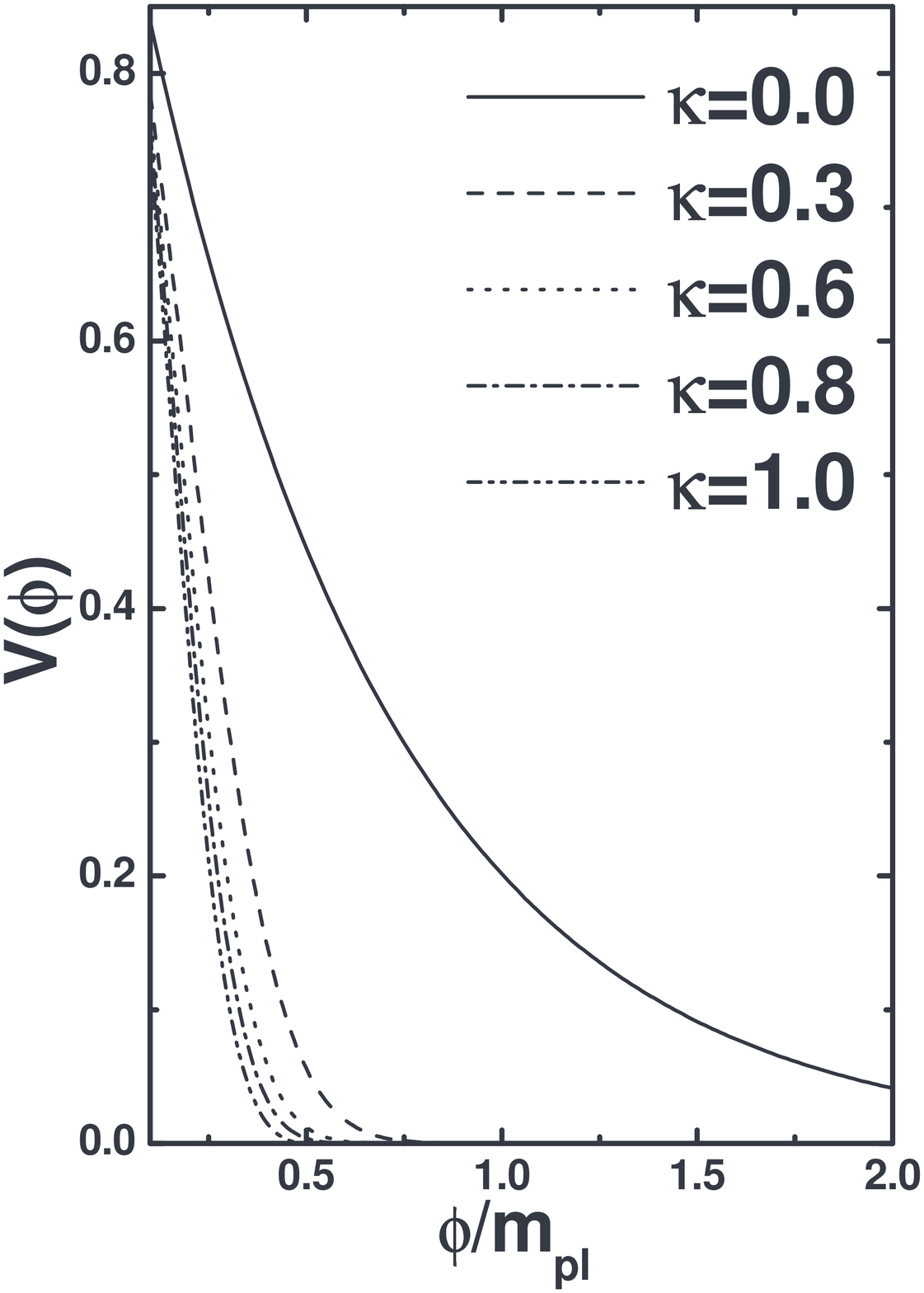,width=2.4truein,height=2.4truein,angle=0}
\psfig{figure=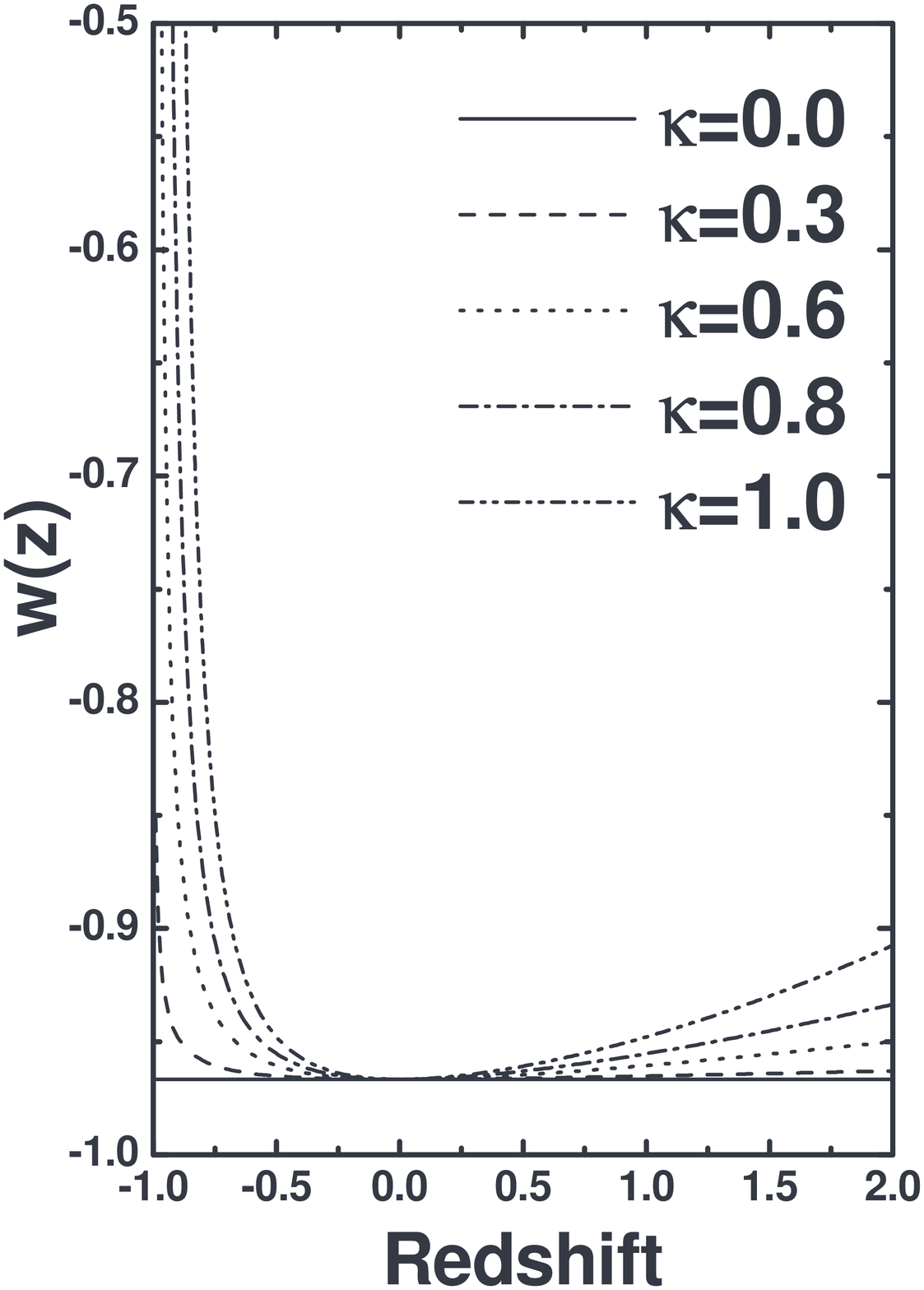,width=2.4truein,height=2.4truein,angle=0}
\psfig{figure=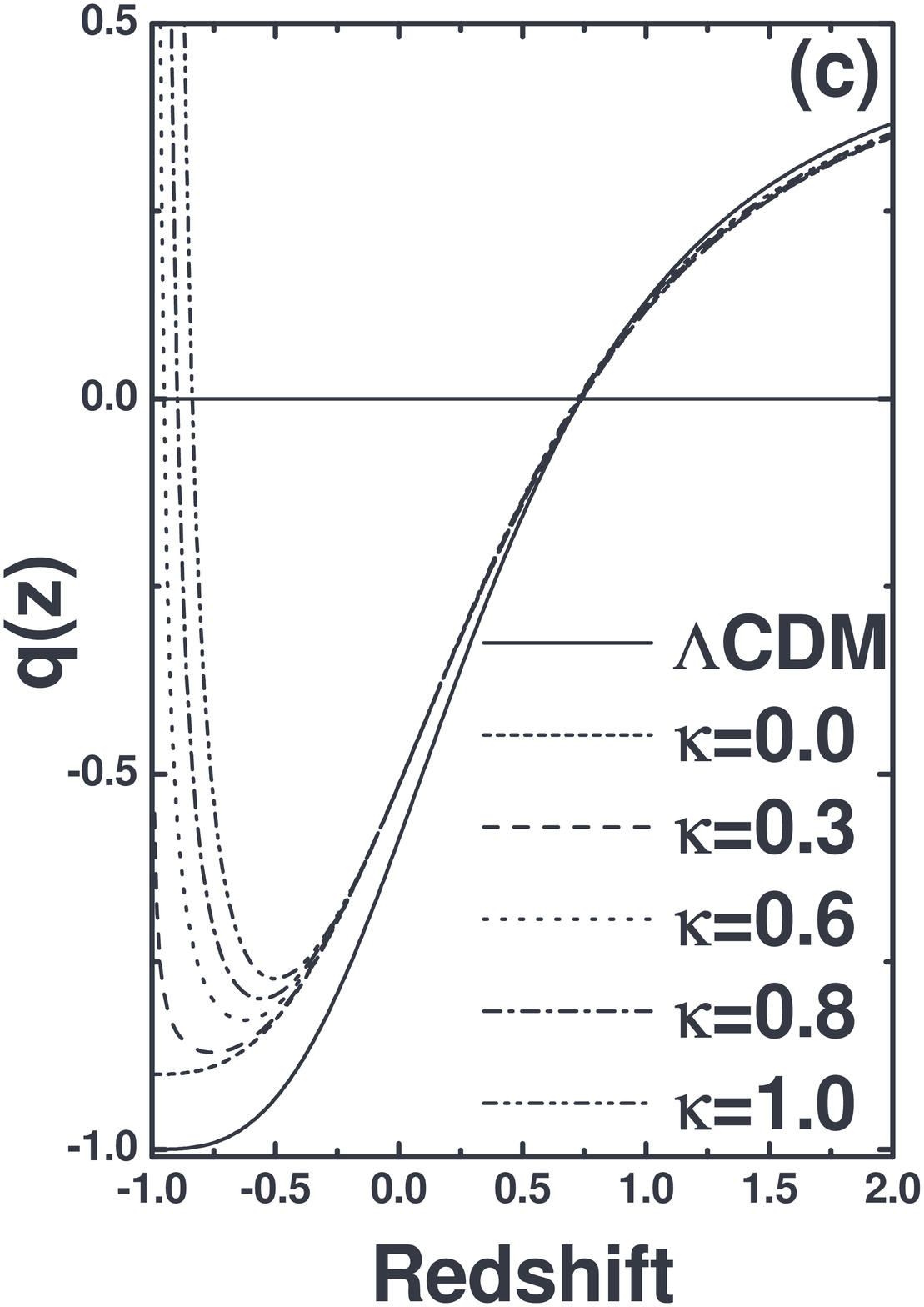,width=2.4truein,height=2.4truein,angle=0}
\hskip 0.01in}
\caption{{\bf{Left:}} The potential $V(\Phi)$ as a function of the field [Eq. (\ref{gpotential})] for some selected values of the index $\kappa$ and $A  = 10^{-1}$. {\bf{Middle:}} The \emph{hybrid} EoS parameter as a function of the redshift. Note that, although behaving as freezing in the past, $w(z)$ becomes thawing as $z$ approaches 0 and will behave as such over the entire future cosmic evolution. {\bf{Right:}} The $q - z$ plane for some selected values of $\kappa$ and $\Omega_{m,0} = 0.27$. Note that for some values of $\kappa \neq 0$ the cosmic acceleration is a transient phenomenon. The $\Lambda$CDM model, whose predicted cosmic acceleration is eternal, is also shown for the sake of comparison.}
\end{figure*}

For the realistic case of a dark matter-dark energy dominated universe (${\cal{L}}_{m} \neq 0$), a direct combination of the above equations provides the following expression for the field $\Phi$ as a function of the scale factor:
\be
\label{phi3}
{\Phi} - {\Phi}_0 = 
\frac{1}{\sqrt{2\sigma}}\int_1^a \frac{a'^{\kappa-1}-a'^{-\kappa-1}}{\sqrt{\left(1 + {\Omega_m(a) \over \Omega_{\Phi}(a)}\right)}}da'\:,
\ee
where $\Omega_m(a)$ and $\Omega_{\Phi}(a)$ stand for the matter and quintessence density parameters, respectively. As one may easily check, the above expression for ${\Phi}(a)$ reduces to $\Phi(a)$ of Eq. (\ref{phi2}) for $\Omega_m = 0$. Note also that, when combined numerically with the  above definitions of $\rho_{\Phi}$ and $p_{\Phi}$, it also provides the potential $V(\Phi)$ for this realistic dark matter/energy scenario, which belongs to the same class of potentials as given in Eq. (\ref{gpotential}) and shown in Figure (1a).

Finally, by integrating out Eq.~(\ref{ansatz}) we obtain the energy density of the field $\Phi$, i.e., 
\be \label{rho_phi1} 
\rho_{\Phi} = \rho_{\Phi,0}a^{-A}\exp({-A\ln_{2\kappa}a})\,, 
\ee 
where $\rho_{\Phi,0}$ is the current value of the energy density. Clearly, in the limit $\kappa \rightarrow0$ Eq. (\ref{rho_phi1}) reduces to an usual power-law, i.e.,
$\rho_{\Phi}(a) \propto a^{-{2A}}$, as expected for ordinary exponential potentials.

\emph{Freezing/Thawing (Hybrid) EoS.}---The EoS parameter for this generalized field is easily derived by combining the conservation equation for $\Phi$ with Eq.(\ref{rho_phi1}), i.e.,
\bea
\label{eq_state}
w(a) = -1 + {A\over 12}(a^{\kappa} + a^{-\kappa})^2\;.
\eea
Figure (1b) shows $w$ as a function of the redshift $z$ for some selected values of the index $\kappa$ (without loss of generality to the subsequent discussions, from now on we particularize our analyses to the case $A = 10^{-1}$). As shown, the scalar field EoS above behaved as freezing over all the past cosmic evolution, is approaching the value $-1$ today [e.g., it is $w(a_0) \simeq -0.96$ for the above value of $A$], will become thawing in the near future and will behave as such over the entire future evolution of the Universe. In particular, at $a \simeq 9^{1/\kappa}$, $w(a)$ crosses the value -1/3, which roughly means the beginning of the future decelerating phase. Clearly, this mixed behavior arises from a competition between the double scale factor terms in Eq. (\ref{eq_state}), which in turn is a direct consequence of the generalized logarithmic function used in our \emph{ansatz} (\ref{ansatz}). Note also that, similarly to the EoS above, all the expressions derived in this paper are symmetric relative to the sign of the index $\kappa$, which means that one may restrict the $\kappa$ interval to $0 \leq \kappa \leq 1$. 

An important aspect worth emphasizing at this point concerns the current observational and theoretical constraints on the behavior of $w(a)$. Although there is so far no concrete observational evidence for a time or redshift-dependence of the dark energy EoS (the current data are fully compatible with the standard $\Lambda$CDM model), some recent analyses using current data from SNe Ia, LSS and CMB have explored possible variations in the $w - a$ plane and indicated a slight preference for a freezing behavior over the thawing one~\cite{krauss,trotta,huterer}. For instance, Ref.~\cite{huterer} uses the Monte Carlo reconstruction formalism to scan a wide range of possibilities for $w(a)$ and find that $\sim 74\%$ are for freezing whereas only $\sim 0.05\%$ are for thawing. %(the remaining possibilities are for models that changes the sign of $dw/d\ln a$ before $z = 0$). 
Similar conclusions are also obtained in Ref.~\cite{trotta} by using the so-called maximum entropy method, where the HST/GOODS SNe Ia data showed $\simeq 1\sigma$ level preference for $w > -1$ at $z \sim 0.5$ with a drift towards $w > -1$ at higher redshifts. These results amount to saying that, if such a preference for freezing potentials persists even after a systematically more homogeneous and statistically more powerful data sets become available, the future of the Universe should be an everlasting acceleration toward a de Sitter phase, which seems to be in conflict with the String/M theories requirements discussed in Ref.~\cite{fischler}. This, however, is not the case in the scenario under discussion here because, differently from pure freezing models, the hybrid EoS given by Eq. (\ref{eq_state}), although freezing in the past (and, therefore, possibly in agreement with the data), will becoming thawing in the future, so that the phenomenon of a transient acceleration may take place. In what follows, and in connection with the results of Ref.~\cite{fischler}, we explore the thawing branch of Eq. (\ref{eq_state}).

\emph{Eternal deceleration.}---Ref.~\cite{fischler} has seriously pointed out a possible conflict between an eternally accelerating universe and our best candidate for a consistent quantum theory of gravity, i.e., String/M theories. The reason for that is because the only known formulation of string theory is in terms of S-matrices which require infinitely separated, noninteracting  in and out states. As is well known, if the universe is dominated by dark energy (for instance, a positive cosmological constant or a freezing potential), it will asymptotically become a de-Sitter space, which has a cosmological event horizon $\Delta_{\rm{h}}$ with physics confined to a finite region. In other words, this means that in eternally accelerating universes there are no isolated states.

In this regard, given the future thawing behavior of the hybrid EoS (\ref{eq_state}), the quintessence scenario discussed here must lead to a transient acceleration. To study this phenomenon, let us consider the deceleration parameter $q = -a\ddot{a}/\dot{a}^{2}$, given by
\begin{widetext}
\begin{equation}
\label{accel_param}
q(z)+1= \frac{\frac{3}{2}\Omega_m(1+z)^3 
+ \frac{A}{8}(1-\Omega_m)(1+z)^{A/2}[2+(1+z)^{2\kappa}+(1+z)^{-2\kappa}]
\exp{\left[-\frac{A}{2}\frac{(1+z)^{-2\kappa}-(1+z)^{2\kappa}}{4\kappa}\right]}}
{\Omega_m (1+z)^3+(1-\Omega_m)(1+z)^{A/2} 
\exp\left[-\frac{A}{2}\frac{(1+z)^{-2\kappa}-(1+z)^{2\kappa}}{4\kappa}\right]}\;,
\end{equation}
\end{widetext} 
and shown in Fig. 1(c) as a function of the redshift for some values of the index $\kappa$ and $\Omega_{m,0} = 0.27$. As can be seen from this figure, for a large interval of values of $\kappa$ the Universe was decelerated in the past, switched to the current accelerating phase at $z_{a} \simeq 1$ and will eventually decelerate again at some $z_{d} < -1$. As expected, this latter transition is becoming more and more delayed as $\kappa \rightarrow 0$. With such a behavior, it is straightforward to show that the cosmological event horizon for this hybrid scenario $\Delta_{\rm{h}} = \int{da/a^2 H(a)}$ diverges, thereby alleviating the potential theoretical and observational conflict discussed above. It is worth emphasizing that, differently from the results of Ref.~\cite{nos06} for a pure thawing EoS, the transient acceleration here is driven by the hybrid EoS of Eq.~(\ref{eq_state}), which behaved as freezing over the entire evolution of the Universe until recently, at $z \simeq 0$.

\begin{figure}[t]
\centerline{
\psfig{figure=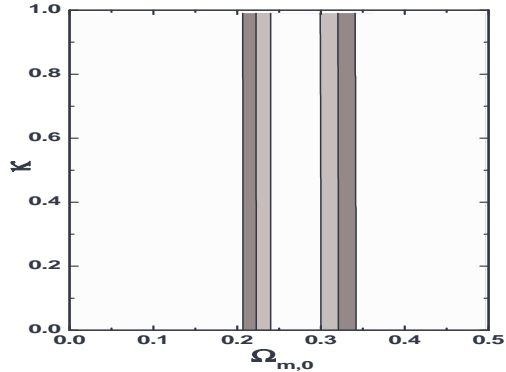,width=2.7truein,height=2.0truein,angle=0} 
\hskip 0.1in} % above
\caption{Confidence contours (68.3\%, 95.4\% and 99.7\%) in the $\Omega_{m,0} - \kappa$ parametric space from a joint analysis involving SNe Ia plus BAO data. Note that no interesting bound can be place on $\kappa$ since the entire interval $0 \leq \kappa \leq 1$ is allowed at $1\sigma$ level. At  $95.4\%$ C.L., we found $\Omega_{m,0} = 0.27 \pm 0.04$.}
\end{figure}

\emph{Observational status and Conclusions.}---In order to place some observational bounds on the $\Omega_m - \kappa$ space, we use a combined sample of 192 SNe Ia events compiled in Ref.~\cite{davis}, which consists of the best quality light-curves SNe Ia of ESSENCE \cite{wv07} and SNLS \cite{Astier06} collaborations and the High-$z$ SNe team \cite{Riess07}, plus the measurement of the baryon acoustic oscillation (BAO) from the Sloan Digital Sky Survey \cite{bao}. Figure (2) shows 68.3\%, 95.4\% and 99.7\% confidence contours in the $\Omega_{m,0} - \kappa$ plane. Note that, similarly to what happens with the time-dependent part of several EoS parameterizations, the current bounds on the index $\kappa$ are weak (with the entire interval $0 \leq \kappa \leq 1$ compatible at 1$\sigma$ level), whereas for the matter density parameter we have found $\Omega_{m,0} = 0.27 \pm 0.02$ at $95.4\%$ (C.L.). We expect, however, that more theoretical effort along with the next generation of experiments dedicated to probe the dark energy EoS as, for instance, the Joint Efficient Dark-energy Investigation (JEDI) \cite{jedi}, the Dark Universe Explore (DUNE) \cite{dune}, among others will have sufficient accuracy to decide which (if pure freezing, thawing or \emph{hybrid}) EoS behavior are preferable from both theoretical and observational viewpoints.

\acknowledgments
JSA and RS are supported by CNPq - Brazil. FCC is supported by FAPESP. Z-HZ's work is supported by the National Science Foundation of China.

\end{document}